# Shear Induced Pressure Determines a Reduction in Polymer Radius

Dave E. Dunstan* and Dalton J. E. Harvie
*Department of Chemical Engineering, University of Melbourne, VIC 3010, Australia.*

davided@unimelb.edu.au, daltonh@unimelb.edu.au

ABSTRACT.

Shear induced particle pressure has been measured and modelled for concentrated suspensions of particles. Importantly, the significance of the shear induced particle pressure has not been recognized in polymer rheology. The shear induced particle pressure results in an inward pressure on the polymer chains resulting in a shear dependent compressive force. The analytical form of the force balance equations that incorporate the effect of shear induced particle pressure predict a reduced polymer blob size and reducing viscosity with increasing shear rate as has been observed experimentally. Power law behavior is found for the viscosity in accord with the general behavior observed for the rheology of concentrated polymer solutions and melts. Differing powers are found for the behavior depending on the concentration regime.

INTRODUCTION.

Polymer dynamics is of wide practical importance and fundamental interest.[1] The models developed to describe polymer chains and their dynamics in flow derive from statistical mechanics and have the capacity to describe polymeric material mechanical properties.[2-6] An elegant connection between molecular properties and the macroscopic behavior has resulted.[1]

A generally accepted assumption of the field is that the polymer chains extend in flow due to the hydrodynamic forces on the chains.[4] Kuhn was the first to propose the dumbbell model that is used in most of the current modelling of polymers in flow in order to directly relate the macroscopic properties of the polymers in flow to the single chain physics.[7] The dumbbell model, in which the polymers are modelled as two beads on an elastic (generally Hookean) FENE spring, is still used in modified forms as it is thought to encapsulate the key physics of the macromolecules under hydrodynamic tension.[8] Physically, a dumbbell experiences two Stokes drags during shear that cause extension and compression as the dumbbells precess in Jeffrey orbits[9], however the constitutive dumbbell models do not incorporate these orbits, instead predict polymer extension.[8] Kuhn also developed a statistical mechanical model to predict the Hookean force law for the ideal chains that acts as a restoring force to counter the hydrodynamic forces.[10] Kuhn initially assumed that the dumbbell could either extend or compress in flow as it rotates around the vorticity axis.

Kuhn and Grun (1942) published the first paper to assume that the chains only extend in simple flow.[11] Essentially, they assumed that only extension occurs and ignored Kuhn's original insight that the chains would undergo both compression and extension in simple flow. By assuming that only extension occurs, the relationship between the reduced shear rate and the end-to-end vector of the chains was calculated. The reduced extension versus shear rate shows a limiting extension at high shear rates.[11]

Cottrell, Merrill and Smith reported the first measurement of light scattering from polymer solutions in shear in 1969 [12]. More recently Link and Springer[13] and then Lee, Solomon and Muller[14] measured light scattering on polymer solutions in Couette flow.[15] Generally, the interpreted deformation is significantly less than the Rouse and Zimm models predict.[16, 17] The observed behaviour may also be interpreted as being due to the orientation of the random ensemble of prolate chains in the flow field without the need to invoke any extension of the chains.[14, 18, 19] The overall shear induced orientation of the prolate chains, in Jeffrey orbits, increases the scattering cross section in the direction perpendicular to the vorticity axis and along the flow direction. This results in the appearance of extension parallel to the flow and compression perpendicular to the flow direction. In the quiescent state the solution appears isotropic due to the random orientation of the prolate chains, then becomes anisotropic in flow via the orientation of the chains. Rheo-optic measurements on dilute solutions of polydiacetylenes in Couette flow show increased projection of the chains in the flow direction, with no deformation of the backbone.[14]

Fluorescently labelled DNA in flow has also been experimentally examined by a number of researchers. The first papers in the field were by Smith, Babcock and Chu [20] and in the same year Le Duc, Haber, Bao and Wirtz who used confocal fluorescence microscopy to directly image labelled DNA in Couette flow.[21] In both works the DNA was visualised using fluorescence microscopy with sliding plates to generate Couette flow and maintain the DNA molecules in the field of view. The DNA is claimed to be representative of random chain polymers in solution. The images show a macroscopic "blob" of several segment lengths that does not appear completely representative of a random chain polymer. Simply, the DNA images are not of a random chain whose conformation is determined by entropy.[20, 21] Furthermore, the resolution of the microscopy method determines that compression is difficult to observe.[20] Larson has written a comprehensive review of the rheology of dilute solutions of flexible polymers focusing on the progress and problems.[22] A considerable component of the review is focused on simulations and modelling the data obtained from DNA. A key conclusion is that the measured deformation is less than expected. It should be noted that DNA does not show the same rheological behaviour as that observed for typical random coil polymers. Typical random coil polymers have conformation that is determined by their entropy and show decreasing viscosity with increasing shear rate and increasing temperature.[23, 24] Calf thymus DNA shows decreasing viscosity with shear rate and increasing viscosity with temperature.[24] A recent study by Bravo-Anaya et al. interprets the observed rheological behaviour as resulting from interacting aggregates of the DNA molecules in flow.[23] The interaction between the DNA molecules is suggested to be driven by H-bonding. In summary, the evidence for polymer chain extension in simple flow at high concentrations above critical overlap is less than compelling. Since Kuhn's original paper, the possibility of compression in Couette flow has not been considered and only extension has been assumed in the field.[4, 6, 8, 25-27] The compression component has been ignored for chains in flow, however, recent experimental evidence has shown chain compression in Couette flow at semi-dilute concentrations.[28-30]

Rheo-optical measurements on synthetic polymers have shown chain orientation in dilute solution and compression at concentrations above critical overlap in the semi-dilute region.[14, 29, 31] These experimental results have prompted a revision of the idea of extension being a universal assumption for polymers in simple planar flow. An alternative approach that assumes compression, allows the measured radius-shear rate behaviour to be predicted, and the power law behaviour observed for polymers in flow to be modelled.[30] Furthermore, using a force balance argument that predicts the shear thinning rheological behaviour, also enables the viscosity-radius relationship to be predicted. The predicted power law behaviour of the viscosity-radius is in close agreement with the experimentally observed behaviour.[30, 32] Interestingly, this shows that the viscosity decreases as the radius decreases in a manner that is physically consistent with the observed behaviour for concentrated random chain polymers.[30] The generally observed decrease in viscosity with increasing temperature for polymer solutions and melts has been attributed to coil compression. The decreasing viscosity with increasing shear rate has been assumed to result from coil extension. This apparent physical inconsistency is resolved if the coils are assumed to compress with increasing shear rate.[32]

It is also worth noting that Frith et al.[33, 34] showed that for sterically stabilized particles with a "soft" stabilizing layer that the stabilizing chains are observed to compress with increasing concentration and shear rate. This is strong evidence that the polymer chains experience a net compressive force in flow and that the idea of a compressive shear induced particle pressure that has not been considered by the field results in a shear dependent compressive force on the chains at higher concentrations.

THEORY.

Batchelor and Green posed the idea of shear induced particle pressure in the 1970s.[35-38] The theory has been developed by a number of researchers since the concept was first posed. Notably, Nott and Brady and a number of key papers by Morris and co-workers have added new insight into the rheology of concentrated suspensions.[39-44] The recent review by Morris gives an elegant overview of the development of the field.[44] The first experimental determination of the shear induced particle pressure was by Bagnold in 1954 who suggested the essential idea of a particle pressure under shear in order to explain his observations.[45] The experimental understanding was further developed by Deboeuf et al. in an elegant work where the suspension was placed in a Couette cell and the change in pressure directly measured as a function of the shear rate.[41] The experimental findings of this work show that the shear induced particle pressure is a linear function of the shear rate and the square of the volume fraction of the suspension.

Particles in solution experience at least two additional compressive normal stresses, or pressures, in addition to the hydrodynamic pressure existing in the solute. The random Brownian motion of the particles causes collisions (or interactions) between particles (in addition to solute interactions which are captured by the hydrodynamic pressure), and on average these interactions create a compressive normal stress on the particle: the osmotic pressure. Particle pressure is the second additional pressure, and is similar to osmotic pressure but instead a result of the shear induced motion of the particles. The particle pressure increases as the collisional frequency increases when the suspension is exposed to a shear stress.[39, 40, 46] The increased collisional frequency results in an increased inward pressure on the particle: the particle pressure. The rheology of dense suspensions is reviewed by Guazzelli and Pouliquen.[47]

Herein we develop a model that incorporates the shear induced particle pressure for polymer blobs in flow to predict how the polymer radius, and hence shear viscosity of the suspension, scales as a function of shear rate. The polymers are treated as spherical, porous, elastic particles at concentrations above critical overlap where the excluded volume terms may be neglected.[2-4] It is also assumed that the chains are not entangled. These assumptions are in accord with the Rouse model that ignores both entanglements and hydrodynamic interactions.[16, 48, 49]

The application of an anisotropic shear stress (Couette flow) causes collisions between the particles that results in a compressional stress being applied to the polymers at a local level. The normal stresses realized in the three directions are given as the diagonal components of the particle stress tensor [40]. In general these components are different in magnitude, however, they all have the same sign, meaning that they all act to compress the particles in suspension.[40] The particle pressure is defined as the average diagonal (isotropic) component of the particle stress tensor. In our analysis we assume that particle compression is a function of this average isotropic stress, or particle pressure.[41, 44] Here the form of the particle interaction pressure developed by Brady and Morris[40] and others[43, 50] for concentrated suspensions means that the particle interaction stress scales as:

$$\sigma_H \sim \eta_n \dot{\gamma} \qquad [1]$$

Here the viscosity, $\eta_n$, is representative of the normal stresses developed in the suspension (proportional to $\phi^2$ to leading order), and $\dot{\gamma}$ is the shear rate.

The shear induced interaction stress on the polymers that compresses the blobs is balanced by the elastic restoring force in each chain. Given that the concentration is assumed to be above critical overlap, the excluded volume and osmotic stresses in the system will be uniform, isotropic and constant (independent of shear stress). The elastic stress resulting from the chain deformation may then be written;

$$\sigma_{el} = -\frac{\partial F_{el}}{\partial V} = \frac{9 k_B T}{8 \pi R_0^2 R} \qquad [2]$$

where $F_{el}$ is the elastic free energy of the chain, $k_B$ Boltzmann's constant, $T$ the absolute temperature, $R_0$ the unperturbed chain radius, $R$ the radius of the chain in flow and $V$ the volume of the chains.

Equating the elastic and interactions stresses at steady state yields:

$$R \sim \frac{k_B T}{\eta_n \dot{\gamma}} \qquad [3]$$

To find the particle radius and shear viscosity we represent the shear and normal viscosities by

$$\eta_s \sim \phi^{n_s} \qquad [4]$$
$$\eta_n \sim \phi^{n_n}$$

where the $n_s$ and $n_n$ are viscosity volume fraction exponents corresponding to shear and normal directions. Substitution of the form of Equations [4] in equation [3], and noting that $\phi \sim R^3$, yields:

$$R \sim \dot{\gamma}^{-\frac{1}{3n_n+1}} \quad [5]$$

And:

$$\eta_s \sim \dot{\gamma}^{-\frac{3n_s}{3n_n+1}} \quad [6]$$

In the simplest case of a semi-dilute solution of polymers, to leading order $\eta_n \sim \phi^2$, the radius scales with shear as $R \sim \dot{\gamma}^{-1/7}$ and the volume fraction by $\phi \sim \dot{\gamma}^{-3/7}$. Adopting a general power law for the shear viscosity of the solution, as per Equation [9], the shear viscosity then scales as $\eta_s \sim \dot{\gamma}^{-\frac{3n_s}{7}}$.

Alternatively, the viscosities appearing in Equation [4] may be assumed to be that of a suspension of particles, with relationships taken from the suspension literation. For example, Morris et al.[43] give:

$$\eta_n = \eta_0 \frac{K_n \bar{\phi}^2}{(1-\bar{\phi})^2}$$

$$\eta_s = \eta_0 \left[ 1 + \frac{2.5 \phi_{max} \bar{\phi}^2}{1-} + \frac{K_s \bar{\phi}^2}{(1-\bar{\phi})^2} \right] \quad [7]$$

Boyer et al.[51] has given the same expressions with different $K_n$ and $K_s$ values. There are other relationships in the literature, but they all have a similar form that captures the experimentally observed trends. The viscosity exponents are plotted in Figure 1. Importantly as $\phi$ approaches the maximum packing limit, both viscosities scale in the same manner.[51] Physically this is reasonable given that both are determined by the interaction frequency between blobs under these conditions. The inset in Figure 1 shows the behavior of the $n_i$ values over the full range of volume fractions and that the values converge in the limit. We can express the relationships of Equation [7] in the form of Equation [4] by calculating effective exponents that vary with volume fraction, using $n_i = \frac{dln(\eta_i)}{dln(\phi)}$ for $i = n, s$. These exponents, as well as the resulting radius and shear viscosity scalings (defined by $\dot{\gamma}^{m_\eta}$ and $\dot{\gamma}^{m_R}$) are plotted in Figure 2 for the $K_n$ and $K_s$ values employed by Miller et al.[52]

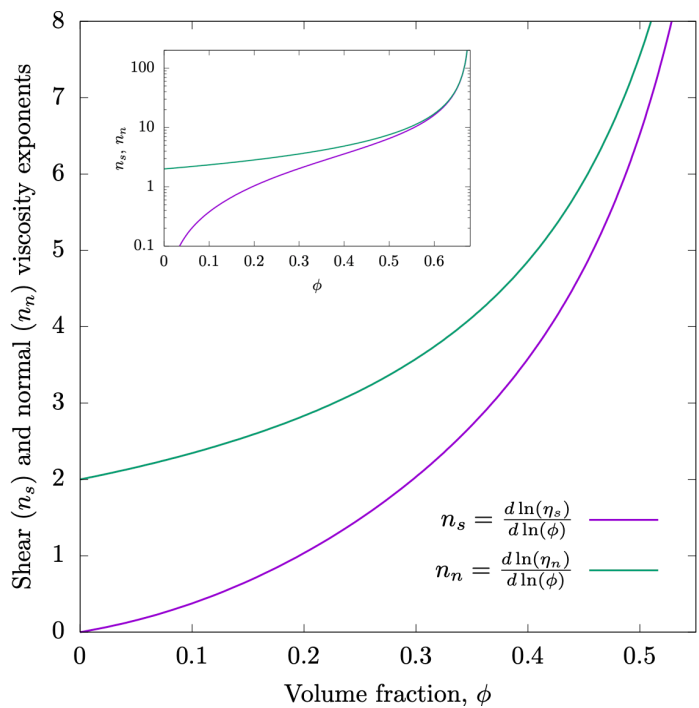

Figure 1. The volume fraction exponents plotted versus the volume fraction using Equation [7] based on values taken from Miller et al.[52] Note that at high volume fraction the values of the normal and shear exponents converge as seen in the inset.

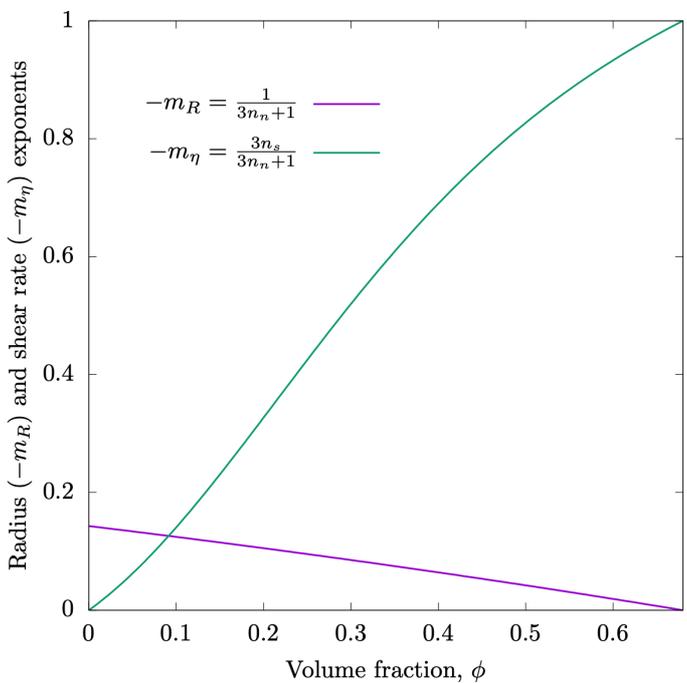

Figure 2. Plot of the shear rate exponents versus volume fraction for the viscosity and radius. Note that the viscosity exponent increases with volume fraction while the radius exponent decreases with volume fraction toward the limiting values of both at the limiting volume fraction.

The form of the power law behavior then becomes a function of the concentration as the values of $n_s$ and $n_n$ both vary as shown in this figure and summarized in Table 1 below.[51]

| Concentration Regime | Exponent Values | Viscosity-Shear rate relationship, $\dot{\gamma}^{m_{eta}}$ | Radius Shear rate relationship, $\dot{\gamma}^{m_R}$ |
|---|---|---|---|
| Dilute | $n_s = 0$, $n_n = 2$ | $\eta_s \sim \dot{\gamma}^{-0} = 1$ | $R \sim \dot{\gamma}^{-1/7}$ |
| Semi-dilute | $n_s = 1$, $n_n = 2$ | $\eta_s \sim \dot{\gamma}^{-3/7}$ | $R \sim \dot{\gamma}^{-1/7}$ |
| Concentrated | $n_s = n_n = n$, | $\eta_s \sim \dot{\gamma}^{\frac{-3n}{3n+1}}$ | $R \sim \dot{\gamma}^{\frac{-1}{3n+1}}$ |
|  | $n \to \infty$ | $\eta_s \sim \dot{\gamma}^{-1}$ | $R \sim 1$ |

Table 1. Results of the predicted power law behavior for the three different concentration regimes. The values of $n_s$ and $n_n$ for the different concentration regimes are taken from Boyer et al. [51]

The current work proposes that polymer compression in flow is a result of the shear induced particle pressure that acts to decrease the chain size with increasing shear rate. Rheo-optical measurements on several random chain polymers at concentrations in the semi-dilute regime have shown compression in flow.[32, 53, 54] Furthermore, power law behavior was observed with the decrease in chain radius with increasing shear rate. The predicted power law range of the radius with shear rate of -1/7 is in reasonable agreement with the experimental value of -0.11 obtained for PMMA.[54] There have been several studies on polymer solutions in shear that indicate the presence of concentration fluctuations due to the imposition of shear.[48, 55] The chain compression predicted in the current study is consistent with the models and experimental evidence collected for polymers in flow using light scattering.[56] In the limit of high concentration, the blobs are highly compressed in the quiescent state and therefore show no further reduction in size with imposed shear.

The model developed here also predicts a power law for the shear thinning behavior that is within the range observed for typical polymer systems.[57, 58] Using the exponent values determined by Boyer et al for the different concentration ranges (see Table 1) enables differing power law behaviours to be determined. The limiting behavior of the shear viscosity being independent of the shear rate in dilute solution to an inverse relationship in concentrated solutions covers the range of power laws observed for polymer solutions. However, further refinement of the model is required to obtain complete agreement. By considering the variation in the exponents of the volume fraction dependence of the shear and normal viscosities in the different concentration regimes, the general behavior of polymers in flow may be modelled. The experimentally determined power law behavior for the viscosity may be fitted and the decrease in the radius with increasing shear rate predicted. Future experimental work on these parameters will enable the validity of the current model to be determined.

CONCLUSIONS.

The shear induced particle pressure results in an overall compressive force on polymer chains in simple Couette flow. By assuming that the chains are permeable blobs and the shear induced pressure is isotropic, the chain size and solution viscosity power law behavour may be predicted. Using the shear induced pressure and elastic restoring force enables the radius-shear rate power law of -1/7 to a constant radius at high concentrations to be predicted while also yielding a power law exponent for the viscosity between 0 and -1 that is within the range of exponents measured for polymer systems.

ACKNOWLEDGEMENT

DD would like to thank Professor George Franks for constructive discussion and listening *ad infinitum.*